\documentclass[aip,jcp,twocolumn,graphicx]{revtex4-1}
\usepackage{todonotes}
\usepackage{xcolor}
\usepackage{amsmath,amsfonts,amssymb}
\usepackage[colorlinks=true,linkcolor=magenta]{hyperref}
\usepackage{multirow}
\usepackage{xspace}
\makeatletter
\DeclareRobustCommand\onedot{\futurelet\@let@token\@onedot}
\newcommand{\@onedot}{\ifx\@let@token.\else.\null\fi\xspace}

\newcommand{\sefour}{Cu$_3$SbSe$_4$\xspace}

\newcommand{\dq}[1]{\lq\lq{}#1\rq\rq{}}
\makeatother
\makeatletter
\def\blfootnote{\xdef\@thefnmark{}\@footnotetext}
\makeatother
\begin{document}

\title{Bonds, bands, and bandgaps in tetrahedrally bonded ternary compounds: the role of group V Lone Pairs}

\author{Dat Do}
\email[]{dodat@msu.edu}
\homepage[ ]{http://www.msu.edu/~dodat}
\author{S. D. Mahanti}
\affiliation{Department of Physics and Astronomy, Michigan State University, East Lansing, MI 48824, USA}

\date{\today}

\begin{abstract}

An interesting class of tetrahedrally coordinated ternary compounds have attracted considerable interest because of their potential as good thermoelectrics. These compounds, denoted as I$_3$-V-VI$_4$, contain three monovalent-I (Cu, Ag), one nominally pentavalent-V (P, As, Sb, Bi), and four hexavalent-VI (S, Se, Te) atoms; and can be visualized as ternary derivatives of the II-VI zincblende or wurtzite semiconductors, obtained by starting from four unit cells of (II-VI) and replacing four type II atoms by three type I and one type V atoms. In trying to understand their electronic structures and transport properties, some fundamental questions arise: whether V atoms are indeed pentavalent and if not how do these compounds become semiconductors, what is the role of V lone pair electrons in the origin of band gaps, and what are the general characteristics of states near the valence band maxima and the conduction band minima. We answer some of these questions using ab initio calculations (density functional methods with both local and nonlocal exchange-correlation potential).

\end{abstract}

\maketitle


\section{Introduction}
Tetrahedrally coordinated semiconductors have played an important role in our understanding of the relationship between coordination, bonding, and band gap. This relationship is easy to understand in monoatomic covalent solids C, Si, Ge, Sn. In these solids, the spatially directed $sp^3$ hybrids form bonding (B) and antibonding (AB) bands.\cite{pauling.1939,phillips.1973} Electrons fill up the bonding bands (valence bands) following the simple Lewis octet rule\cite{lewis_atom_1916,langmuir_arrangement_1919}, the antibonding bands (condution bands) are empty, and there exists a band gap (E$_g$) between the valence band maximum and the conduction band minimum. There is a decrease in the splitting between the bonding and anti-bonding bands due to increase in the lattice constant as one goes from C to Sn, with a concurrent reduction in the band widths. The net effect is a decrease in $E_g$ which is nearly zero for Sn. Although this qualitative relationship between bonds and band gap is easy to understand, the actual value of $E_g$ depends on other subtleties like the details of band dispersion, spin orbit interaction etc. 

The inter-relationship between bonds, bands and band gaps discussed above can be extended to tetrahedrally bonded III-V and II-VI binary compounds.\cite{pauling.1939, phillips.1973} Here again the octet rule plays the dominant role. However, the splitting between the valence and the conduction bands and therefore the band gap depends on other parameters, namely the differences in electron affinity (EA) and ionization energy (IE) of the two components. The bonds now pick up both covalent and ionic characters. The classic works of Pauling\cite{pauling.1939} and Phillips\cite{phillips.1973} on relating the ionicity and covalency of a particular bond to the above parameters (Pauling) and to the dielectric properties (Phillips) have given a fundamental understanding of the bonding and structure of these binary compounds. But these theories do not explicitly address the relationship between band gaps and the nature of bonds. (Phillips' ionicity scale does depend indirectly on the band gap through the dielectric constant). However, the general trend seen in covalent semiconductors, i.e. the decrease in $E_g$ with increasing lattice constant (bond length) is also seen in binary tetrahedrally bonded semiconductors. For example, $E_g = 6.2$~eV in BN (lattice constant $a$=3.61~\AA) and 0.23~eV in InSb ($a$=6.48~\AA). Although this tendency is quite general, the actual value of the band gap depends on the details of the band dispersion. Progress in electronic structure theories (density functional theory (DFT), both local and non-local, GW, and others) have helped us in addressing the relationship between bonding, band structure and band gap in these compounds.\cite{martin.2004}

Extending the above ideas to tetrahedrally coordinated ternary compounds becomes challenging because of the competition between the natures of different bonds. A simple but helpful way of looking at the ternary compounds is to consider them as derivatives of a tetrahedrally coordinated binary compounds, following Grimm and Sommerfeld's rule.\cite{grimm_sommerfeld_rule} One can take multiple unit cells of a ternary system, fix (say) the anions, and replace the cations by a combination of two cations keeping the total cation valence constant. A classic example is the chalcopyrite structure (CuInSe$_2$). One starts from ZnSe with zincblend structure, doubles the unit cell to (ZnSe)$_2$, keeps the Se sublattice intact, and replaces two Zn (divalent -- II) atoms by Cu (monovalent -- I) and In (trivalent -- III). Another example is to triple the unit cell and replace the three divalent cations by two monovalent and one tetravalent cation and get I$_2$-IV-VI$_3$, such as Cu$_2$GeSe$_3$. The system that we will be concerned with here is the class of I$_3$-V-VI$_4$ compounds, where the unit cell is quadruple of ZnSe, and 4 divalent cations are replaced by three monovalent (Cu, Ag) and one pentavalent (P, As, Sb, Bi) cation. Our focus here is to understand the nature the band structure and its relation to the inter-atomic bonds, critically examining the valency of the nominally pentavalent cation ($ns^2np^3$ shells of V, $n=3,4,5,6$) and the role of its lone pair ($ns^2$) in the band-gap formation. 

It is known that the lone pairs play an important role in the structural distortions and the formation of local dipole moments in several physical systems (GeTe, SnTe, PbTe, PbTiO$_3$, PbZrO$_3$, BiVO$_4$, etc).\cite{waghmare_lone_2003,stoltzfus_structure_2007} Sidgwick and Powell\cite{sidgwick_bakerian_1940} in their discussions on the foundations of \emph{Valence Shell Electron Pair Repulsion} theory argued that bonding pairs and lone pairs are of equal importance in the structure formation and these electrons distributed themselves to minimize inter-electron repulsion. Later Gillipsie and Nyholm\cite{gillespie_inorganic_1957,gmespie_electron-pair_1970} argued that repulsive forces between lone pairs and other electrons were stronger. The effect of lone pairs on the atomic structure of inorganic compounds is well known. In fact, cation-centered lone pairs (often with Pb$^{2+}$ as the central cation, but also Sn$^{2+}$ and Bi$^{3+}$) have been found to be important in understanding the origin of off-centered displacement of ions and the resulting electric dipoles.\cite{waghmare_lone_2003,stoltzfus_structure_2007,baettig_anti_2007} These off-centered displacements and associated low-energy structural dynamics\cite{nielsen_lone_thermal_2013} (soft phonons and large Gruneissen parameters) have played important roles in ferro- and piezo-electric materials, multiferroics,  thermoelectrics and  nonlinear optical materials.

In spite of extensive studies trying to understand the role of lone pairs on the atomic structure and lattice vibrations, the subtle role of the interplay between the lone pairs and the local geometry on the band structure and the band-gap formation has not been explored in great depth in the class I$_3$-V-VI$_4$ of compounds. The main aim of this work is to explore this question and point out that in compounds where the group V atom is tetrahedrally coordinated, its lone pairs indeed play an important role not only in opening up a gap (making the system semiconducting) but also in controlling the nature of low-energy electronic excitations. We also point out how the simple Lewis octet rule breaks down when the lone pairs take an active part in the band-gap formation.

The arrangement of the paper is as follows: In section \ref{sec.struct} we describe the structure of tetrahedrally-coordinated ternary compounds I$_3$-V-VI$_4$ containing group V atoms (P, As, Sb, Bi). In section \ref{sect.atomic.en} we discuss the atomic energy levels of the constituent atoms to help us in understanding the results of the ab initio band structure calculations. Section \ref{sec.method} discusses briefly some theoretical and technical details of our calculations and section~\ref{sec.result} gives our results.  Finally we present a summary of our major findings in section~\ref{sec.sum}.
	
\section{\label{sec.struct}Structure of tetrahedrally coordinated I$_3$-V-VI$_4$ compounds}

\begin{figure}
\includegraphics[width=\columnwidth]{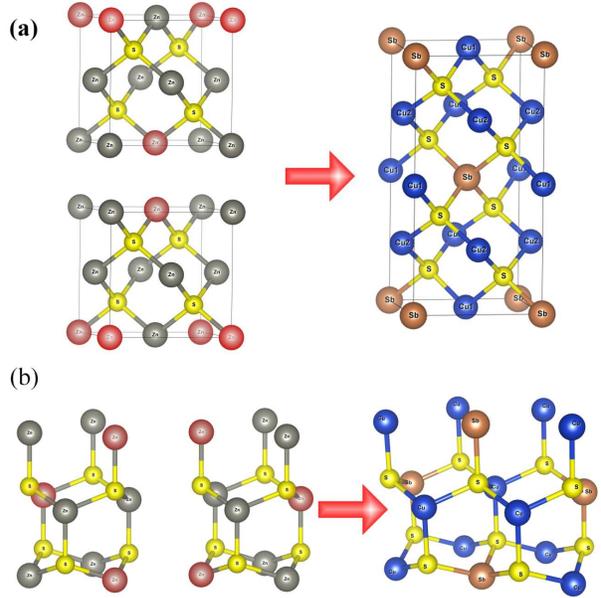}
\caption{\label{fig.cryst} This figure will appear in color in print and online. Crystal structure of Famatinite formed by doubling the Zinc blend unit cell  (top panel) and Enargite formed by doubling the Wurtzite (hexagonal) unit cell (bottom panel). Yellow represents S (Se, Te), gray represents Zn which are replaced by Cu/Ag (blue) and red represents Zn which are replaced by Sb (P, As, or Bi)}
\end{figure}

\begin{table*}
\caption{\label{tab.cryst}Summary of crystal structures of Cu$_3$AB$_4$}
\begin{tabular}{| l | p{.2\textwidth} | p{.2\textwidth} | p{.2\textwidth} | p{.2\textwidth}|}
\hline\hline
B\textbackslash A&P&As&Sb&Bi\\
\hline
S& En\newline $d = 2.06^{exp.}\cite{garin_crystal_1972,pfitzner_refinement_2002}$\newline\hspace*{3.5ex}2.11$^{GGA}$ & Fa/En($T_c=580$)\cite{datahandbook}\newline $d = 2.20^{exp.}$\cite{pfitzner_system_2004} \newline \hspace*{3.5ex} 2.31$^{GGA}$& Fa\newline $d = 2.38^{exp.}$\cite{pfitzner_refinement_2002}\newline\hspace*{3.5ex}2.48$^{GGA}$ & Fa\newline d = 2.65$^{GGA}$\\\hline
Se& En\newline $d = 2.29-2.36^{exp.}$\cite{garin_crystal_1972}\newline\hspace*{3.5ex}$2.29-2.32^{GGA}$& Fa\newline $d = 2.50^{GGA}$ & Fa\newline$d = 2.54^{exp.}$\cite{pfitzner94}\newline\hspace*{3.5ex}$2.65^{GGA}$ & Fa\newline d = 2.78$^{GGA}$\\\hline
Te& $\times$ & ? & ? & $\times$\\
\hline\hline
\end{tabular}
\newline\newline En = Enargite, Fa = Famatinite, $\times$ = does not exist, ? = unknown, $d$ = nearest neighbor distance between atoms A and B (unit Angstrom). GGA indicates that parameters were relaxed using GGA approximation in the current work.
\end{table*}

We consider the family of I$_3$-V-VI$_4$ compounds where (I) is Cu and Ag, (V) is P, As, Sb or Bi, and (VI) is S, Se or Te. While most Ag-compounds are not found in the literature, Cu-compounds usually crystallize in the Famatinite structure (Fa) and some in the Enargite structure (En). Famatinite and Enargite structures may be considered as derivatives of the well known Zinc blende and Wurtzite structures respectively. As shown in Fig.~\ref{fig.cryst}, Fa and En can be obtained by doubling the ZnS unit cell (Zincblende or Wurtzite respectively) and replacing four Zn atoms by three I's and one V. This keeps the cation valence count intact (4 divalent Zn vis-\`a-vis 3 monovalent Ag/Cu and 1 nominally pentavalent P/As/Sb/Bi). In our calculations, due to different space groups (body centered tetragonal for Fa and simple orthorhombic for En), the Fa unit cell contains one formula unit while En unit cell has two formula units. This difference gives rise to different Brillouin zones and a difference in the number of bands for a given wave vector \textbf{k}, namely, En has twice as many bands  as Fa.

In both the structures, all the atoms are tetrahedrally coordinated to four nearest neighbors, in which I and V atoms bond to four VI atoms while VI atoms bond to three I and one V. Table~\ref{tab.cryst} presents a summary of crystal types of I$_3$-V-VI$_4$ compounds and the bond lengths between V and VI. As the constituent atoms go from smaller to larger radii, the bond lengths increase accordingly. We will show later that the change in bond lengths and the band gaps are intimately related. 

\section{\label{sect.atomic.en}Atomic energy levels of constituents and some schematic studies.}

\begin{table*}
\caption{\label{tab.atomiclevel} Atomic energies (given in eV) of the $s$ and $p$ valence electrons for different constituents and for the monovalent Cu, the energy of Cu $d$-level is included.\cite{harrison_table}}
\begin{tabular}{lllllllll}
\hline\hline
&P&As&Sb&Bi&Cu&S&Se&Te\\
\hline
E$_s$&-19.22&-18.92&-16.03&-15.19&-7.7&-24.02&-22.86&-19.12\\
E$_p$&-9.54&-8.98&-8.14&-7.79&--&-11.6&-10.68&-9.54\\
E$_d$&--&--&--&--&-20.26&--&--&--\\
\hline\hline
\end{tabular}
\end{table*}

\begin{figure}
\includegraphics[width=\columnwidth]{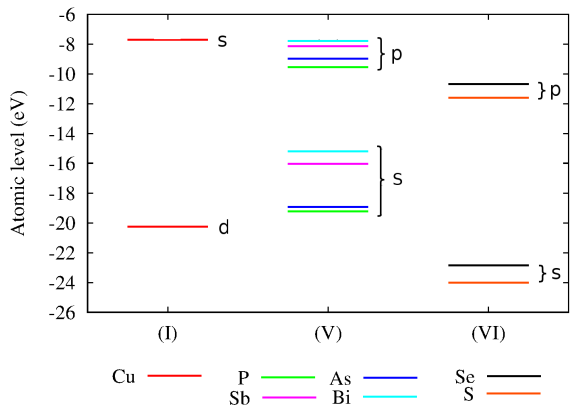}
\caption{\label{fig.atom_level}This figure will appear in color in print and online. Visualization of the atomic levels of the constituent atoms.}
\end{figure}

In analyzing bonding and antibonding natures of different bonds and bands, it is helpful to know the energies of atomic levels of the constituent atoms. In table~\ref{tab.atomiclevel}, we give the atomic energies of the $s$ and $p$ valence electrons for different constituents and for Cu we also give the energy of the $d$-level. Positions of these atomic levels are shown in Fig.~\ref{fig.atom_level}. It is clear that Cu-$d$ and S/Se-$s$ level are the lowest, which makes both S/Se-$s$ and Cu-$d$ states stable and fully occupied, Cu-$d$ states locate near the Fermi level.  The energies of the $s$ and $p$ levels of V-atoms increase in going from P to Bi.

\begin{figure}
\includegraphics[width=\columnwidth]{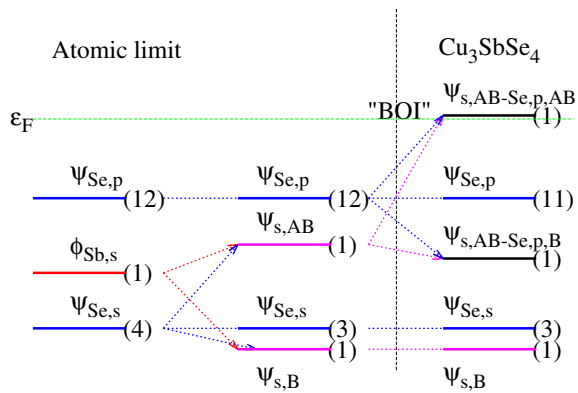}
\caption{\label{fig.se4.level.diagram}This figure will appear in color in print and online. Simple bonding-antibonding scheme in \sefour showing energy levels in the atomic limit (left column) intermediate bonding-antibonding states (middle column) and bonding-antibonding states in crystal (right column). (Cu $d$ and Sb-$p$ are not shown for simplicity). In the blankets are the numbers of states.}
\end{figure}

In our earlier work on Cu$_3$SbSe$_4$,\cite{do_cusbse.2012} we proposed a simplified bonding-antibonding model for states of different atoms as shown in Fig.~\ref{fig.se4.level.diagram}. In this picture, $s$-level of Cu and $p$-states of Sb have highest energy and donate all their electrons to the others. The $s$-states of Se can be treated as filled core states and the $s$-states of Cu as empty states to start. The $p$-states of Se anions and the $s$-state of Sb (lone pairs) strongly interact and form the bonding and anti-bonding bands, giving rise to an unoccupied band near the Fermi level (hereinafter we call this band the "band of interest" or \dq{BOI}). The position of the BOI is very sensitive to the approximation scheme applied to the exchange-correlation potential and the distance between the Sb atom and its four Se neighbors. The lone pair states of Sb are somewhat schizophrenic, they not only bond with the $s$-states of the surrounding Se but also like to bond with one of the properly symmetrized combination of the $p$ states of those Se$_4$ cluster. This latter bonding is so strong that it splits off one band from the top of the Se $p$-bands, giving rise to a band gap. The $d$-bands of Cu are quite narrow and lie below but near the Fermi level. Their precise role in the band structure depends on the particular type of atoms forming the compound. Our aim in this paper is to extract some of the generic (universal) role of the V lone pairs and show that the proposed bonding-antibonding picture for Cu$_3$SbSe$_4$ is rather universal and applicable to the other tetrahedral compounds containing V-VI$_4$ clusters. Since the nearest neighbor distances increase from P to Bi one should expect competing effects between the shrinking of bonding-antibonding separation and the narrowing of the bandwidth on the formation of the band gap. This competing relation is more complicated in these ternary compounds because they have several types of atoms with different atomic energy levels and different natures of bonds. Particularly, to understand the role of $d$ states of I, we investigate the case with Cu replaced by an alkali atom, Na, which doesn't have the $d$-shells to interact with other bands.

\section{\label{sec.method}Methods of electronic structure calculations}
\subsection{\label{subsec.dft}Local/Semi-local DFT}

Firstly, we tried the standard LDA and GGA since they have been extensively used to study the electronic structure of compounds. One, however, is aware of their well-known shortcomings in dealing with semiconductors and ordinary insulators (band insulators).\cite{defects} The major problem is the underestimation of the band gap.\cite{defects} In some extreme cases, gapped system are wrongly predicted to be metallic within LDA/GGA.\cite{do_cusbse.2012,do_fe2val.2011}  There have been many attempts to improve LDA/GGA and several methods have been proposed, including semilocal approximations: PBE for solid (PBE-sol)\cite{pbesol}, modified Becke-Johnson potential by Tran and Blaha (mBJ),\cite{mbj} LDA/GGA+U.\cite{anisimov91,ldau_anisimov97}

LDA/GGA+U method was proposed several decades ago by introducing a correction due to Coulomb interaction of opposite-spin localized electrons when they occupied the same orbital. This correction is a mean field approximation of Hubbard-type correction, in which energy levels become:
\begin{equation}
\epsilon_i = \epsilon^{LDA}_i+U(\frac12-n_i)
\end{equation}
where $n_i$ is the occupancy of energy level $\epsilon_i$. GGA+U has been successful in predicting the semiconducting/insulating ground states of many systems, including but not limited to the famous Mott-insulators.\cite{ldau_anisimov97}

\subsection{\label{subsec.hybrid}Hybrid Functional}

Among several improvements beyond LDA/GGA, hybrid functionals\cite{hybrid} are being increasingly used. In this scheme, to account for the discontinuity of the exchange potential, which is missed in the (semi)local theories, a part of the exchange term  in LDA/GGA is replaced by an exact Hartree-Fock term. The formalism developed by Heyd, Scuseria and Ernzernhof in 2003 and then in 2006 (HSE06),\cite{hse06:1,hse06:2,hse06:3} has become a desired method of choice to study electronic structure for many authors. In this scheme, the exchange-correlation energy is given by
\begin{align}
E_{xc}^{LDAh}&=E_{x}^{LDAh}+E_{c}^{LDA}\\
E_{x}^{LDAh}&=\alpha{}E_x^{HF,SR}(\omega)+(1-\alpha)E_x^{PBE,SR}(\omega)\notag\\
&+E_x^{PBE,LR}(\omega)
\end{align}
in which LR and SR denote long-range and short-range parts respectively, $\alpha$ and $\omega$ are the mixing and screening parameters accordingly. $\omega=0$ mean no screening at all and $\omega\rightarrow\infty$ mean complete screening, or in other words, HSE06 in the limit of infinite $\omega$ falls back to GGA. The values of $\alpha$=1/4 and $\omega$=0.2 are considered optimal values which give reasonable results for many systems.\cite{hse06:1,hse06:2,hse06:3}

\subsection{\label{subsec.tech}Technical Details}
Calculations were done using the projector-augmented wave (PAW)
method\cite{bloch94, kresse99} as implemented in the VASP code.\cite{vasp1,vasp2,vasp3} Plane-wave energy cutoff of 400 eV and an energy convergence criterion (between two successive self-consistent cycles) of $10^{-4}$ eV (total energy/unit cell) were set. We use Perdew-Burke-Ernzerhof (PBE) exchange-correlation functional\cite{pbe} for GGA. For GGA+U calculations we apply the formalism proposed by Dudarev et al.\cite{dudarev98}, where the on-site repulsion and exchange are incorporated through a parameter $U_{eff}=U-J$, where $U$ and $J$ are the Coulomb and exchange parameters. Monkhorst-Pack \textbf{k}-mesh of $12\times12\times12$ for Fa and $12\times14\times10$ for En were used. The density of states (DOS) is obtained using bigger \textbf{k}-mesh. For HSE06 type, due to extensive computational requirement, smaller \textbf{k}-meshes are used, within acceptable uncertainty.

\section{\label{sec.result}Results and discussions}

\begin{figure}
\includegraphics[width=\columnwidth]{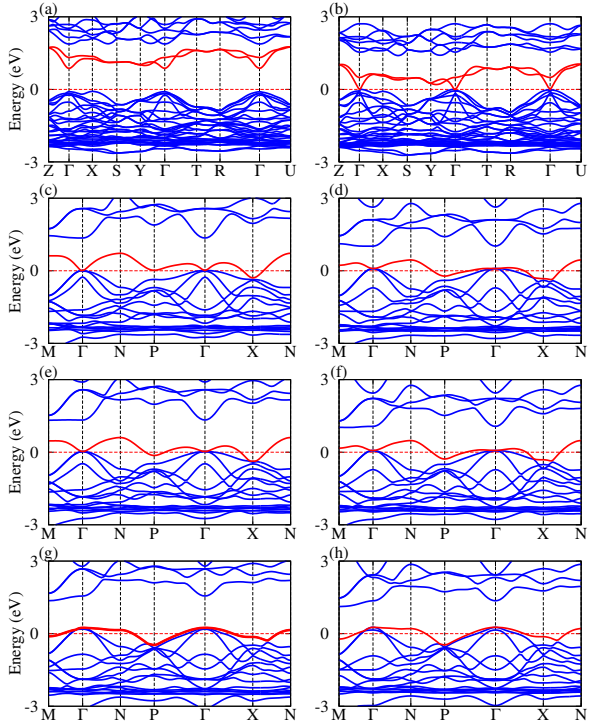}
\caption{\label{fig.gga.band}This figure will appear in color in print and online. GGA bandstructure of (a) Cu$_3$PS$_4$ (b) Cu$_3$PSe$_4$ (c) Cu$_3$AsS$_4$ (d) Cu$_3$AsSe$_4$ (e) Cu$_3$SbS$_4$ (f) Cu$_3$SbSe$_4$ (g) Cu$_3$BiS$_4$ and (h) Cu$_3$BiSe$_4$.}
\end{figure}

To understand the crucial role played by the lone pairs of the group V atoms in the bonding-antibonding scheme and whether the underlying mechanism for the formation of band gap proposed in Cu$_3$SbSe$_4$ can be applied to other members of the I$_3$-V-VI$_4$ family, we first carried out GGA calculations the whole class; Cu$_3$(P,As,Sb,Bi)(S,Se)$_4$. In Fig.~\ref{fig.gga.band} we show the band structures of these compounds, focusing near the Fermi level, in which the BOI's are marked as red thick lines. The GGA calculations show that all the compounds have similar features in their band structures which resemble the simple picture of the bonding-antibonding scheme discussed above. The difference between En and Fa band structures is that for each \textbf{k} the former has twice as much number of bands as the latter. As a result En has two BOI's. Both compounds have Cu-$d$ states occupied right below the Fermi level and the lowest partially unoccupied band is the BOI. The Cu-$d$ states mix with the Se-$p$ states. Most systems, like Cu$_3$SbSe$_4$, within GGA, are semimetals with pseudogaps near the Fermi energy. For these systems, the Cu-$d$ states mix with the BOI at the $\Gamma$ point of the Brillouin zone, giving rise to a three-fold degenerate band, while the next band has Sb-$s$ and Se-$p$ (or V-$s$ and VI-$p$ for the others) characters as the BOI. The only exception is Cu$_3$PS$_4$ where there is a band gap between BOI and the rest of the bands below, which are occupied. It is predicted to be a semiconductor even with GGA level of approximation. The reason which makes Cu$_3$PS$_4$ unique is that it has the shortest V-VI nearest neighbor distance as listed in Table~\ref{tab.cryst}.  We also find that the overlap between BOI and valence bands increases when going from P to Bi as well as from S to Se. 

\begin{figure}
\begin{minipage}{.8\columnwidth}
\hspace*{-20em}(a)\\
\includegraphics[width=\columnwidth]{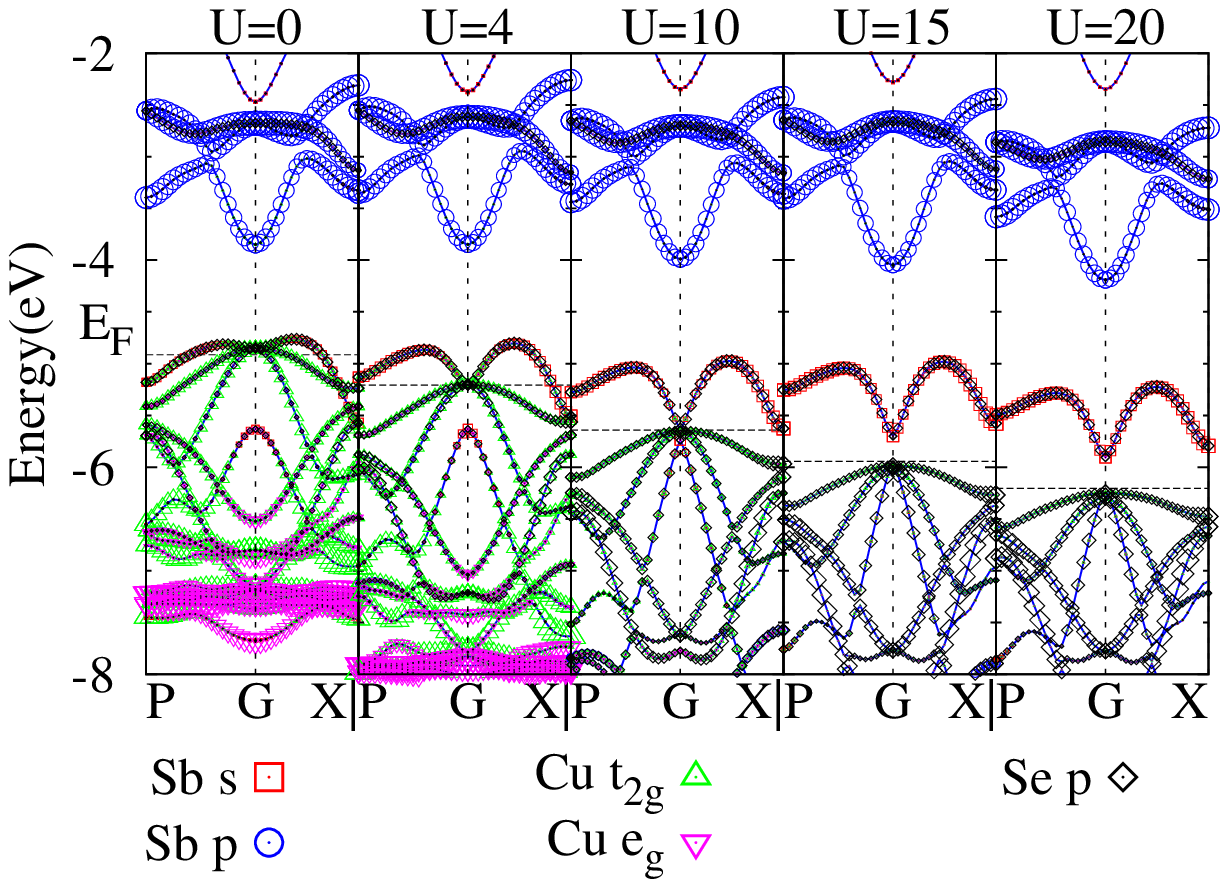}
\end{minipage}\\
\begin{minipage}{.8\columnwidth}
\hspace*{-20em}(b)\\
\includegraphics[width=\columnwidth]{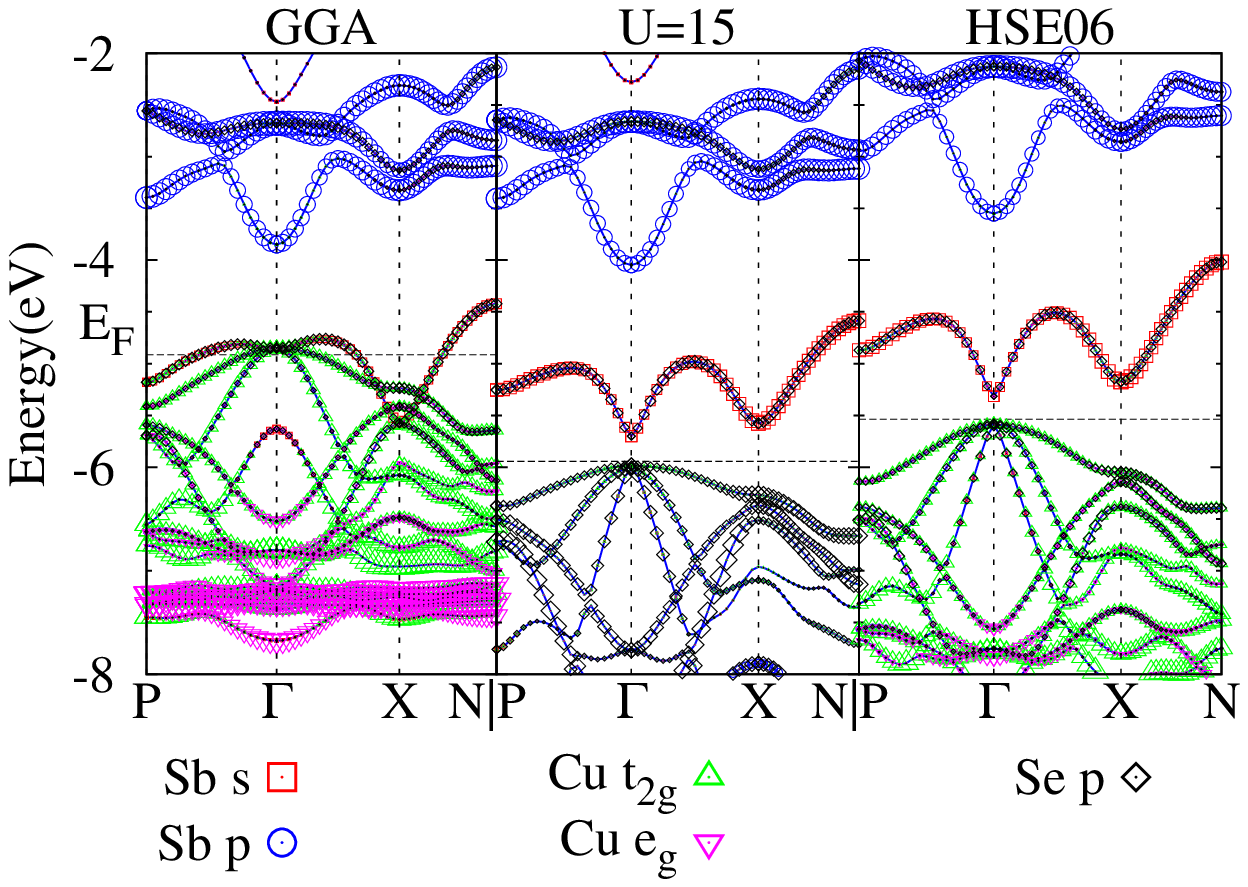}
\end{minipage}
\caption{\label{fig.se4_uhse_band}This figure will appear in color in print and online. (a) Effect of U on band structure and (b) comparison between GGA+U and HSE06 band structures. The energy is given in the absolute scale. }
\end{figure}

In our earlier work\cite{do_cusbse.2012}, to overcome the GGA’s failure to predict the band gap correctly we used the HSE06 scheme which correctly produced a band gap of $\sim${0.3} eV in Cu$_3$SbSe$_4$ (in agreement with experiment\cite{Skoug.doping.2011,nakanishi69,datahandbook}). A similar scheme was used by \citet{yang11.newptype} and also gave a band gap, the value of the gap, however, is $\sim$0.4~eV which is a little larger than our value. We also noted that one can get almost identical results within GGA+U (local theory) but using an unphysically large value of U ($\sim${15}~eV) for the Cu-$d$ orbitals. Indeed, in spite of the unphysical meaning large U ($\sim$15~eV), GGA+U could reproduce almost identical band structures near the Fermi level  as those obtained from HSE06. As seen in Fig.~\ref{fig.se4_uhse_band}a, increasing U does two things, it lowers the occupied $d$-bands of Cu, which reduces the hybridization between the Cu $d$ and the rest of the $s$, $p$ bands (the amount of $d$-character in BOI at the $\Gamma$ point decreases) and eventually, for a sufficiently large value of U, a gap opens up between the BOI and the rest of the occupied valence bands. The value of U needed to open a gap is too large ($>$10 eV) in comparison to the typical value used in literature\cite{Blaha.ldau.cu.2005}. Similar effects are observed in HSE06 calculations\cite{do_cusbse.2012} as one increases the strength of the nonlocal exchange by increasing the value of the mixing parameter $\alpha$, the $d$-character of the BOI at the $\Gamma$ point decreases and the band gap opens up. We found that with U=15~eV, GGA+U calculation could reproduce reasonably well HSE06 band structure near the Fermi energy (see Fig.~\ref{fig.se4_uhse_band}b).

One subtle difference between GGA+U and HSE06 is that the Cu-$d$ bands are pushed down much further in the former. It can be easily observed in Fig.~\ref{fig.se4_uhse_band}. In Fig.~\ref{fig.se4_uhse_band}a, as U increases, the mixing between Cu $d$ and Se $p$ decreases, and as a result, the $d$ characters in the conduction bands near Fermi level get weaker and eventually disappear at large U. The difference in the position of $d$-levels is easily seen when comparing band structures obtained using GGA, GGA+U and HSE06. This suggests that if HSE06 gives the correct band structure, one should detect a Cu-$d$ peak closer to Fermi level in XPS. 

Despite of the difference in the position of $d$-states, the band structures near the Fermi level are quite similar in GGA+U and HSE06. Since the low-energy physics of materials (the materials studied in this work in particular) depend mostly on the electronic structure in the vicinity of the Fermi level, GGA+U (even with unphysical U) can give some good insight in different compounds, without consuming much computational resources as needed for HSE06 calculations. Hence, GGA+U is easier to implement when one studies the nature of defect states in these compounds because in this case one will have to use large supercells with number of atoms ($\sim${64} or larger).

\begin{figure}
\includegraphics[width=\columnwidth]{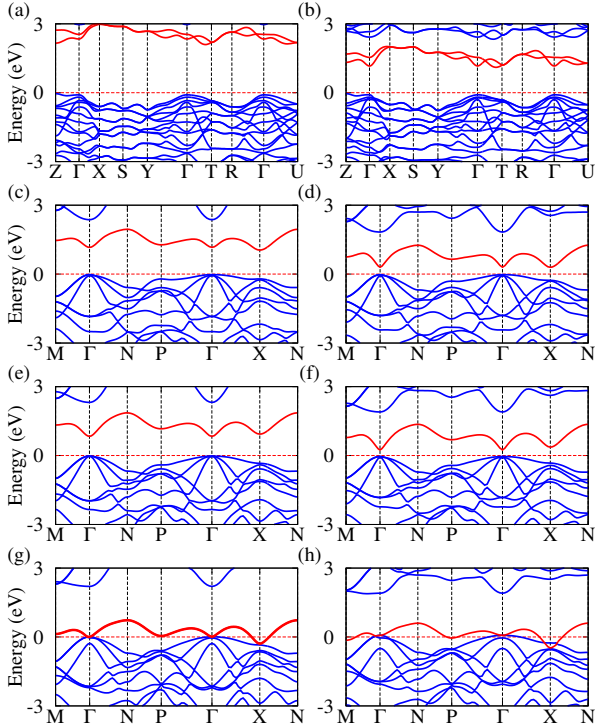}
\caption{\label{fig.ggau.band}This figure will appear in color in print and online. GGA+U bandstructure of (a) Cu$_3$PS$_4$ (b) Cu$_3$PSe$_4$ (c) Cu$_3$AsS$_4$ (d) Cu$_3$AsSe$_4$ (e) Cu$_3$SbS$_4$ (f) Cu$_3$SbSe$_4$ (g) Cu$_3$BiS$_4$ and (h) Cu$_3$BiSe$_4$.}
\end{figure}

With the above rationale, we use GGA+U calculations to investigate the general role of the lone pair of V in the band gap formation in the entire class of I$_3$-V-VI$_4$ compounds. Whenever we have some questions for a particular system, we can use HSE06 calculations to compare. In Fig.~\ref{fig.ggau.band}, we give the GGA+U band structures of (a) Cu$_3$PS$_4$, (b) Cu$_3$PSe$_4$, (c) Cu$_3$AsS$_4$, (d) Cu$_3$AsSe$_4$, (e) Cu$_3$SbS$_4$, (f) Cu$_3$SbSe$_4$, (g) Cu$_3$BiS$_4$ and (h) Cu$_3$BiSe$_4$. We see that a band gap between BOI and the occupied valence bands opens up, except in Bi compounds. For those compounds where band gap opens up, the highest occupied bands are multiply degenerate, which makes these systems excellent  $p$-type thermoelectric, just like Cu$_3$SbSe$_4$.\cite{do_cusbse.2012}

\begin{figure}
\includegraphics[width=\columnwidth]{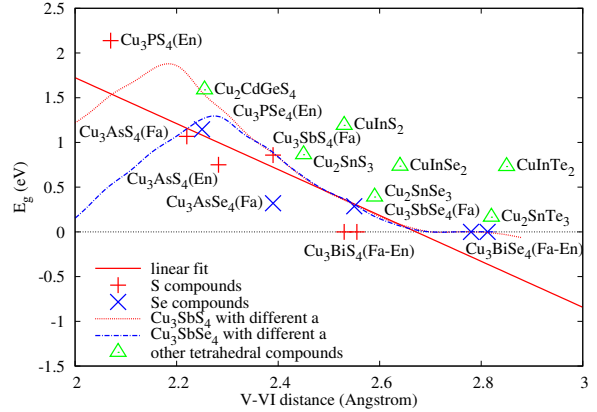}
\caption{\label{fig.d_eg_ldau}This figure will appear in color in print and online. Relation between band gap and the distance between V and VI elements.}
\end{figure}

To see how the band structures depend on the crystal structure, for some compounds, we carried out calculations for both Fa and En structures. In Fig.~\ref{fig.d_eg_ldau} we plot the magnitude of the band gap as a function or average nearest neighbor (nn) distance between V and VI elements and between the $sp$ elements for the other ternary compounds involving lone-pair atoms. 

Let us first analyze the trend in I$_3$-V-VI$_4$ compounds. One can see that irrespective of the crystal structure (En or Fa), the band gap decreases with the increase of the V-VI distance ($d_{V-VI}$). The band gap decreases almost linearly with the $d_{V-VI}$ up to $d\approx2.6$\AA{} as represented by the red solid line in Fig.~\ref{fig.d_eg_ldau}. With larger $d_{V-VI}$, the gap disappears. With other constituents being the same, compounds with S have smaller $d_{V-VI}$ than those with Se, and thus the band gaps are larger in the former. However, for the same $d_{V-VI}$, Se-compounds have larger band gap. For instance, the band gap of Cu$_3$SbSe$_4$ is $\sim$0.4~eV while Cu$_3$BiS$_4$ is gapless. One exception is the case of Cu$_3$SbS$_4$ and Cu$_3$AsSe$_4$. Both of them have $d_{V-VI}\approx2.4$\AA{}  but the former has a larger band gap. 

To explain these features, one may attempt to look at the atomic energy levels of the constituents (given in table~\ref{tab.atomiclevel} and Fig.~\ref{fig.atom_level}). The difference between S and Se is that the atomic levels of the latter is higher than those of the former, the energy difference for $s$ level is $\sim$1.16~eV and for $p$ is $\sim$0.92~eV. These energy differences may account for the band gap difference between S- and Se-compounds. On the other hand, when going from As to Sb, there is a big jump in the atomic levels (Fig.~\ref{fig.atom_level}). This change in the atomic levels between As and Sb may compensate for the difference in the atomic levels between S and Se, making the band gap of Cu$_3$AsSe$_4$ smaller than that of Cu$_3$SbS$_4$. 

To check this argument, we vary $d_{V-VI}$ in Cu$_3$SbS$_4$ and Cu$_3$SbSe$_4$ by changing the lattice parameter, keeping the relative ionic positions intact. The results are presented in Fig.~\ref{fig.d_eg_ldau} as the purple dotted (Cu$_3$SbS$_4$) and cyan dot-dashed (Cu$_3$SbSe$_4$) curves. Interestingly, for $d_{V-VI}>2.3$\AA, two curves are almost identical and follow the general trend which we discuss above; band gap decreases when the $d_{V-VI}$ increases. For small distances, two curves split by about 1~eV and the trend turns over; the band gap now decreases with decreasing distance. This change indicates that for short distances, the increasing of the band width overcomes the splitting of bonding-antibonding band, reducing the band gap. The turning points are different in two curves, it occurs at larger distance in Se-compound. These results do not appear to support the arguments we made about the relation between the band gap and atomic levels. 

\begin{table}
\caption{\label{tab.same_a}Band-gap comparison between different V elements at the same lattice parameter as Cu$_3$SbSe$_4$ with and without relaxation.}
\begin{tabular}{|l|l|l|l|}
\hline\hline
\multirow{2}{*}{Element}&\multicolumn{2}{|c|}{w/ relaxation}&\multirow{2}{*}{w/o relaxation}\\
\cline{2-3}
&$a$(\AA)&$E_g$(eV)&\\
\hline
P&2.29&0.92&0\\
As&2.42&0.29&0\\
Sb&2.56&0.28&0.28\\
Bi&2.68&0&0\\
\hline\hline
\end{tabular}
\end{table}

In an attempt to resolve this puzzle, we calculate the band structures for Cu$_3$XSe$_4$, where X is P, As, Sb or Bi, using the same lattice parameter as that of Cu$_3$SbSe$_4$ for two cases: (1) with the same positions of atoms (hence, the same $d_{V-VI}$) and (2) with the ionic positions relaxed. The results are shown in table~\ref{tab.same_a}. It is interesting to note that, without relaxation, only Sb-compound has a positive band gap whereas other compounds have overlap between BOI and the valence bands. However, when the atomic positions are allowed to relax, we go back to the general trend where the $d_{V-VI}$ increases as one goes from P to Bi, and as a result, the band gap decreases accordingly. Thus, different constituents affect the local geometry of tetrahedrally coordinated compounds differently, giving different values of the band gaps.

Similar effects are also observed in other tetrahedrally coordinated compounds such as I-III-VI$_2$, I$_2$-IV-VI$_3$. The values of the band gaps (within GGA+U, with U=15~eV) for some systems are shown as green triangle in Fig.~\ref{fig.d_eg_ldau}. These materials, in general, have larger band gaps than the I$_3$-V-VI$_4$ compounds. Detailed studies of these compounds will be discussed in another paper. 

\begin{figure}
\centering
\includegraphics[width=1\columnwidth]{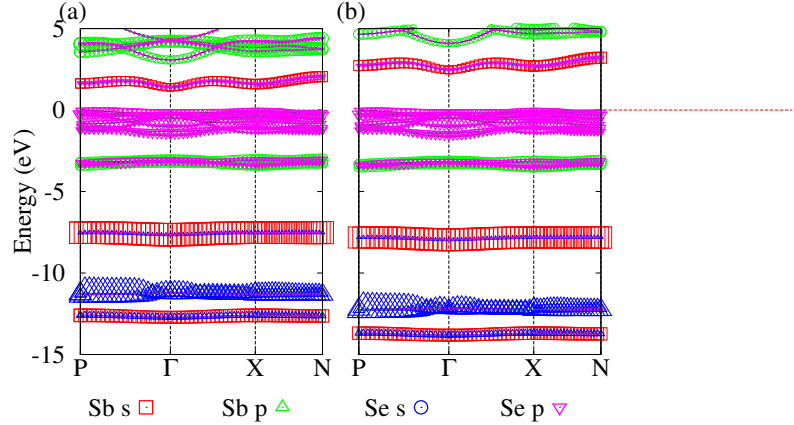}\\
\caption{\label{fig.Na_band}This figure will appear in color in print and online. Band structure of Na3SbSe4 obtained using (a) GGA and (b) HSE06 with fatband representation showing atomic orbital associated with energy levels.}
\end{figure}

To verify that the underlying physics of the band-gap formation is dominated by the interaction between the lone pairs of V and the $p$-states of VI and less so by the hybridization with the $d$-states of Cu, we carried out calculations for an artificial compound Na$_3$SbSe$_4$. The structural parameters, since it is, to the best of our knowledge, not known in the literature, are initially chosen to be the same as Cu$_3$SbSe$_4$ but then allowed to fully relax. In Fig.~\ref{fig.Na_band}, we give the band structures of Na$_3$SbSe$_4$ obtained using GGA (U=0) and HSE06 calculations. GGA already opens up a band gap of $\sim$1.5~eV similar to Cu$_3$PS$_4$ and HSE06 increases the band gap to $\sim$2.5~eV. The bands are rather narrow compared to the Cu compounds because the lattice constant for the Na compound is 6.29~\AA{} compared to 5.73~\AA{} for the Cu compound.  Also the distance between Sb and nn Se decreases from 2.65~\AA{} in the Cu compound to 2.49~\AA{} in the Na compound. This leads to a stronger interaction between V and VI atoms leading to a larger band gap in Na$_3$SbSe$_4$. Another interesting feature shows up, clearly due to the absence of $d$-levels, we find that other nondegenerate bands split off from the Se-$p$ bands but pushed below. This band has strong Sb-$p$ character indicating a strong mixing between the Sb-$p$ conduction bands and the Se-$p$ valence bands. A closer examination of the band structure of the Cu compounds also shows this feature. 

\begin{figure}
\centering
\includegraphics[width=\columnwidth]{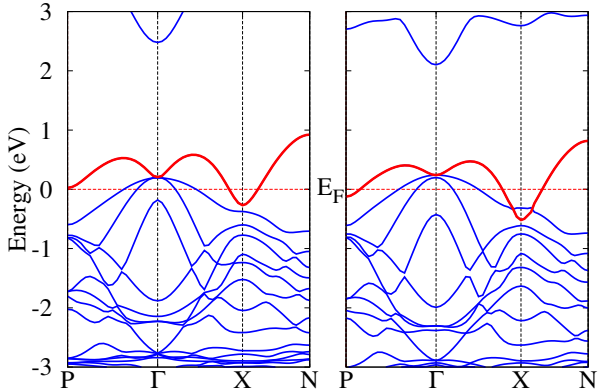}
\caption{\label{fig.hse.band}This figure will appear in color in print and online. HSE06 bandstructure of (a) Cu$_3$BiS$_4$ and (b) Cu$_3$BiSe$_4$.}
\end{figure}

Finally we would like to address the band structure of the two compounds containing Bi, Cu$_3$BiS$_4$ and Cu$_3$BiSe$_4$. Both GGA (Fig.~\ref{fig.gga.band}g,h) and GGA+U (Fig.\ref{fig.ggau.band}g,h) did not give a gap. We carried out HSE06 calculations for these two compounds and the results are given in Fig.~\ref{fig.hse.band}. Again we do not see any gap opening up and both these compounds are semimetals and the band structures are similar to those obtained from GGA+U. We thus conclude that Cu$_3$BiS$_4$ and Cu$_3$BiSe$_4$ are semimetals.

\section{\label{sec.sum}Summary and Conclusion}

In summary, we have studied the electronic structures of valence compensated ternary compounds given by the formula I$_3$-V-VI$_4$ and its relation to the local bonds as well as to the role of V lone-pair electrons. We found that V lone pair electrons strongly interact with the surrounding VI-$p$ orbitals, giving rise to the special BOI (single band in Famatinite and two bands in Enargite) which play the role of the lowest of conduction bands. V lone pairs are directly involved in the band gap formation. On the other hand, the filled $d$-shells of Cu contribute indirectly in the band gap, only through changing the bonding nature and bond length between V and VI atoms. By comparing the electronic structure of a large number of ternary I$_3$-V-VI$_4$ systems using different methods (local and non-local), we bring out the importance of the non-local exchange in gap opening. We also show that GGA+U (with a properly chosen value of U, say 15~eV in this work) is still useful in surveying a wide range of materials, giving adequate information to understand the physics of materials, particularly, at low energy. Our work also suggests that one can easily tune the band gap to a desirable value by choosing proper constituent elements. This work gives a foundation for further study of tetrahedrally bonded ternary materials, for example to investigate the electronic structure of defects. 

\begin{acknowledgements}
This work was supported by the Center for Revolutionary Materials for Solid State Energy Conversion, an Energy Frontier Research Center funded by the U.S. Department of Energy, Office of Science, Office of Basic Energy Sciences under Award Number DE-SC0001054.

The calculations were done using computational resource provided by National Energy Research Scientific Computing Center(NERSC) and Michigan State University (MSU), Institution for Cyber Enabled Research (ICER), and High Performance Computer Center (HPCC). 
\end{acknowledgements}
\bibliography{reference,octet_rule_ref}   

\begin{thebibliography}{41}
\expandafter\ifx\csname natexlab\endcsname\relax\def\natexlab#1{#1}\fi
\expandafter\ifx\csname bibnamefont\endcsname\relax
  \def\bibnamefont#1{#1}\fi
\expandafter\ifx\csname bibfnamefont\endcsname\relax
  \def\bibfnamefont#1{#1}\fi
\expandafter\ifx\csname citenamefont\endcsname\relax
  \def\citenamefont#1{#1}\fi
\expandafter\ifx\csname url\endcsname\relax
  \def\url#1{\texttt{#1}}\fi
\expandafter\ifx\csname urlprefix\endcsname\relax\def\urlprefix{URL }\fi
\providecommand{\bibinfo}[2]{#2}
\providecommand{\eprint}[2][]{\url{#2}}

\bibitem[{\citenamefont{Pauling}(1939)}]{pauling.1939}
\bibinfo{author}{\bibfnamefont{L.}~\bibnamefont{Pauling}},
  \emph{\bibinfo{title}{The Nature of Chemical Bond and the Structure of
  Molecules and Crystals}} (\bibinfo{publisher}{Cornell University Press},
  \bibinfo{address}{New York}, \bibinfo{year}{1939}).

\bibitem[{\citenamefont{Phillips}(1973)}]{phillips.1973}
\bibinfo{author}{\bibfnamefont{J.~C.} \bibnamefont{Phillips}},
  \emph{\bibinfo{title}{Bonds and Bands in Semiconductors}}
  (\bibinfo{publisher}{Academic Press}, \bibinfo{address}{New York and London},
  \bibinfo{year}{1973}).

\bibitem[{\citenamefont{Lewis}(1916)}]{lewis_atom_1916}
\bibinfo{author}{\bibfnamefont{G.~N.} \bibnamefont{Lewis}},
  \bibinfo{journal}{J. Am. Chem. Soc.} \textbf{\bibinfo{volume}{38}},
  \bibinfo{pages}{762} (\bibinfo{year}{1916}), ISSN \bibinfo{issn}{0002-7863}.

\bibitem[{\citenamefont{Langmuir}(1919)}]{langmuir_arrangement_1919}
\bibinfo{author}{\bibfnamefont{I.}~\bibnamefont{Langmuir}},
  \bibinfo{journal}{J. Am. Chem. Soc.} \textbf{\bibinfo{volume}{41}},
  \bibinfo{pages}{868} (\bibinfo{year}{1919}), ISSN \bibinfo{issn}{0002-7863}.

\bibitem[{\citenamefont{Martin}(2004)}]{martin.2004}
\bibinfo{author}{\bibfnamefont{R.~M.} \bibnamefont{Martin}},
  \emph{\bibinfo{title}{Electronic Structure, Basic Theory and Practical
  Method}} (\bibinfo{publisher}{Cambridge University Press},
  \bibinfo{year}{2004}).

\bibitem[{\citenamefont{Grimm and Sommerfeld}(1926)}]{grimm_sommerfeld_rule}
\bibinfo{author}{\bibfnamefont{H.}~\bibnamefont{Grimm}} \bibnamefont{and}
  \bibinfo{author}{\bibfnamefont{A.}~\bibnamefont{Sommerfeld}},
  \bibinfo{journal}{Zeitschrift f\"ur Physik} \textbf{\bibinfo{volume}{36}},
  \bibinfo{pages}{36} (\bibinfo{year}{1926}), ISSN \bibinfo{issn}{0044-3328}.

\bibitem[{\citenamefont{Waghmare et~al.}(2003)\citenamefont{Waghmare, Spaldin,
  Kandpal, and Seshadri}}]{waghmare_lone_2003}
\bibinfo{author}{\bibfnamefont{U.~V.} \bibnamefont{Waghmare}},
  \bibinfo{author}{\bibfnamefont{N.~A.} \bibnamefont{Spaldin}},
  \bibinfo{author}{\bibfnamefont{H.~C.} \bibnamefont{Kandpal}},
  \bibnamefont{and} \bibinfo{author}{\bibfnamefont{R.}~\bibnamefont{Seshadri}},
  \bibinfo{journal}{Phys. Rev. B} \textbf{\bibinfo{volume}{67}},
  \bibinfo{pages}{125111} (\bibinfo{year}{2003}).

\bibitem[{\citenamefont{Stoltzfus et~al.}(2007)\citenamefont{Stoltzfus,
  Woodward, Seshadri, Klepeis, and Bursten}}]{stoltzfus_structure_2007}
\bibinfo{author}{\bibfnamefont{M.~W.} \bibnamefont{Stoltzfus}},
  \bibinfo{author}{\bibfnamefont{P.~M.} \bibnamefont{Woodward}},
  \bibinfo{author}{\bibfnamefont{R.}~\bibnamefont{Seshadri}},
  \bibinfo{author}{\bibfnamefont{J.-H.} \bibnamefont{Klepeis}},
  \bibnamefont{and} \bibinfo{author}{\bibfnamefont{B.}~\bibnamefont{Bursten}},
  \bibinfo{journal}{Inorg. Chem.} \textbf{\bibinfo{volume}{46}},
  \bibinfo{pages}{3839} (\bibinfo{year}{2007}), ISSN \bibinfo{issn}{0020-1669}.

\bibitem[{\citenamefont{Sidgwick and Powell}(1940)}]{sidgwick_bakerian_1940}
\bibinfo{author}{\bibfnamefont{N.~V.} \bibnamefont{Sidgwick}} \bibnamefont{and}
  \bibinfo{author}{\bibfnamefont{H.~M.} \bibnamefont{Powell}},
  \bibinfo{journal}{Proc. R. Soc. Lond. A} \textbf{\bibinfo{volume}{176}},
  \bibinfo{pages}{153} (\bibinfo{year}{1940}), ISSN \bibinfo{issn}{1364-5021,
  1471-2946}.

\bibitem[{\citenamefont{Gillespie and Nyholm}(1957)}]{gillespie_inorganic_1957}
\bibinfo{author}{\bibfnamefont{R.~J.} \bibnamefont{Gillespie}}
  \bibnamefont{and} \bibinfo{author}{\bibfnamefont{R.~S.}
  \bibnamefont{Nyholm}}, \textbf{\bibinfo{volume}{11}}, \bibinfo{pages}{339}
  (\bibinfo{year}{1957}), ISSN \bibinfo{issn}{0009-2681}.

\bibitem[{\citenamefont{Gmespie}(1970)}]{gmespie_electron-pair_1970}
\bibinfo{author}{\bibfnamefont{R.~J.} \bibnamefont{Gmespie}},
  \bibinfo{journal}{J. Chem. Educ.} \textbf{\bibinfo{volume}{47}},
  \bibinfo{pages}{18} (\bibinfo{year}{1970}), ISSN \bibinfo{issn}{0021-9584}.

\bibitem[{\citenamefont{Baettig et~al.}(2007)\citenamefont{Baettig, Seshadri,
  and Spaldin}}]{baettig_anti_2007}
\bibinfo{author}{\bibfnamefont{P.}~\bibnamefont{Baettig}},
  \bibinfo{author}{\bibfnamefont{R.}~\bibnamefont{Seshadri}}, \bibnamefont{and}
  \bibinfo{author}{\bibfnamefont{N.~A.} \bibnamefont{Spaldin}},
  \bibinfo{journal}{Journal of the American Chemical Society}
  \textbf{\bibinfo{volume}{129}}, \bibinfo{pages}{9854} (\bibinfo{year}{2007}).

\bibitem[{\citenamefont{Nielsen et~al.}(2013)\citenamefont{Nielsen, Ozolins,
  and Heremans}}]{nielsen_lone_thermal_2013}
\bibinfo{author}{\bibfnamefont{M.~D.} \bibnamefont{Nielsen}},
  \bibinfo{author}{\bibfnamefont{V.}~\bibnamefont{Ozolins}}, \bibnamefont{and}
  \bibinfo{author}{\bibfnamefont{J.~P.} \bibnamefont{Heremans}},
  \bibinfo{journal}{Energy Environ. Sci.} \textbf{\bibinfo{volume}{6}},
  \bibinfo{pages}{570} (\bibinfo{year}{2013}).

\bibitem[{\citenamefont{Garin and Parthé}(1972)}]{garin_crystal_1972}
\bibinfo{author}{\bibfnamefont{J.}~\bibnamefont{Garin}} \bibnamefont{and}
  \bibinfo{author}{\bibfnamefont{E.}~\bibnamefont{Parthé}},
  \textbf{\bibinfo{volume}{28}}, \bibinfo{pages}{3672} (\bibinfo{year}{1972}),
  ISSN \bibinfo{issn}{05677408}.

\bibitem[{\citenamefont{Pfitzner and Reiser}(2002)}]{pfitzner_refinement_2002}
\bibinfo{author}{\bibfnamefont{A.}~\bibnamefont{Pfitzner}} \bibnamefont{and}
  \bibinfo{author}{\bibfnamefont{S.}~\bibnamefont{Reiser}},
  \textbf{\bibinfo{volume}{217}}, \bibinfo{pages}{51} (\bibinfo{year}{2002}),
  ISSN \bibinfo{issn}{0044-2968}.

\bibitem[{\citenamefont{Madelung}(2004)}]{datahandbook}
\bibinfo{author}{\bibfnamefont{O.}~\bibnamefont{Madelung}},
  \emph{\bibinfo{title}{Semiconductors: Data Handbook}}
  (\bibinfo{publisher}{Springer}, \bibinfo{year}{2004}), \bibinfo{edition}{3rd}
  ed.

\bibitem[{\citenamefont{Pfitzner and Bernert}(2004)}]{pfitzner_system_2004}
\bibinfo{author}{\bibfnamefont{A.}~\bibnamefont{Pfitzner}} \bibnamefont{and}
  \bibinfo{author}{\bibfnamefont{T.}~\bibnamefont{Bernert}},
  \textbf{\bibinfo{volume}{219}}, \bibinfo{pages}{20} (\bibinfo{year}{2004}),
  ISSN \bibinfo{issn}{0044-2968}.

\bibitem[{\citenamefont{Pfitzner}(1994)}]{pfitzner94}
\bibinfo{author}{\bibfnamefont{A.}~\bibnamefont{Pfitzner}},
  \bibinfo{journal}{Z. Kristallogr.} \textbf{\bibinfo{volume}{209}},
  \bibinfo{pages}{685} (\bibinfo{year}{1994}).

\bibitem[{\citenamefont{Harrison}(2004)}]{harrison_table}
\bibinfo{author}{\bibfnamefont{W.~A.} \bibnamefont{Harrison}},
  \emph{\bibinfo{title}{Elementary electronic Structure}}
  (\bibinfo{publisher}{World Scientific}, \bibinfo{address}{Singapore},
  \bibinfo{year}{2004}).

\bibitem[{\citenamefont{Do et~al.}(2012)\citenamefont{Do, Ozolins, Mahanti,
  Lee, Zhang, and Wolverton}}]{do_cusbse.2012}
\bibinfo{author}{\bibfnamefont{D.}~\bibnamefont{Do}},
  \bibinfo{author}{\bibfnamefont{V.}~\bibnamefont{Ozolins}},
  \bibinfo{author}{\bibfnamefont{S.~D.} \bibnamefont{Mahanti}},
  \bibinfo{author}{\bibfnamefont{M.-S.} \bibnamefont{Lee}},
  \bibinfo{author}{\bibfnamefont{Y.}~\bibnamefont{Zhang}}, \bibnamefont{and}
  \bibinfo{author}{\bibfnamefont{C.}~\bibnamefont{Wolverton}},
  \bibinfo{journal}{Journal of Physics: Condensed Matter}
  \textbf{\bibinfo{volume}{24}}, \bibinfo{pages}{415502}
  (\bibinfo{year}{2012}).

\bibitem[{\citenamefont{Nieminen}(2006)}]{defects}
\bibinfo{author}{\bibfnamefont{R.~M.} \bibnamefont{Nieminen}},
  \emph{\bibinfo{title}{Topics in Applied Physics: Theory of defects in
  semiconductors}} (\bibinfo{publisher}{Springer}, \bibinfo{year}{2006}), vol.
  \bibinfo{volume}{104}, pp. \bibinfo{pages}{36--40}.

\bibitem[{\citenamefont{Do et~al.}(2011)\citenamefont{Do, Lee, and
  Mahanti}}]{do_fe2val.2011}
\bibinfo{author}{\bibfnamefont{D.}~\bibnamefont{Do}},
  \bibinfo{author}{\bibfnamefont{M.-S.} \bibnamefont{Lee}}, \bibnamefont{and}
  \bibinfo{author}{\bibfnamefont{S.~D.} \bibnamefont{Mahanti}},
  \bibinfo{journal}{Phys. Rev. B} \textbf{\bibinfo{volume}{84}},
  \bibinfo{pages}{125104} (\bibinfo{year}{2011}).

\bibitem[{\citenamefont{Perdew et~al.}(2008)\citenamefont{Perdew, Ruzsinszky,
  Csonka, Vydrov, Scuseria, Constantin, Zhou, and Burke}}]{pbesol}
\bibinfo{author}{\bibfnamefont{J.~P.} \bibnamefont{Perdew}},
  \bibinfo{author}{\bibfnamefont{A.}~\bibnamefont{Ruzsinszky}},
  \bibinfo{author}{\bibfnamefont{G.~I.} \bibnamefont{Csonka}},
  \bibinfo{author}{\bibfnamefont{O.~A.} \bibnamefont{Vydrov}},
  \bibinfo{author}{\bibfnamefont{G.~E.} \bibnamefont{Scuseria}},
  \bibinfo{author}{\bibfnamefont{L.~A.} \bibnamefont{Constantin}},
  \bibinfo{author}{\bibfnamefont{X.}~\bibnamefont{Zhou}}, \bibnamefont{and}
  \bibinfo{author}{\bibfnamefont{K.}~\bibnamefont{Burke}},
  \bibinfo{journal}{Phys. Rev. Lett.} \textbf{\bibinfo{volume}{100}},
  \bibinfo{pages}{136406} (\bibinfo{year}{2008}).

\bibitem[{\citenamefont{Tran and Blaha}(2009)}]{mbj}
\bibinfo{author}{\bibfnamefont{F.}~\bibnamefont{Tran}} \bibnamefont{and}
  \bibinfo{author}{\bibfnamefont{P.}~\bibnamefont{Blaha}},
  \bibinfo{journal}{Phys. Rev. Lett.} \textbf{\bibinfo{volume}{102}},
  \bibinfo{pages}{226401} (\bibinfo{year}{2009}).

\bibitem[{\citenamefont{Anisimov and Gunnarsson}(1991)}]{anisimov91}
\bibinfo{author}{\bibfnamefont{V.~I.} \bibnamefont{Anisimov}} \bibnamefont{and}
  \bibinfo{author}{\bibfnamefont{O.}~\bibnamefont{Gunnarsson}},
  \bibinfo{journal}{Phys. Rev. B} \textbf{\bibinfo{volume}{43}},
  \bibinfo{pages}{7570} (\bibinfo{year}{1991}).

\bibitem[{\citenamefont{Anisimov et~al.}(1997)\citenamefont{Anisimov,
  Aryasetiawan, and Lichtenstein}}]{ldau_anisimov97}
\bibinfo{author}{\bibfnamefont{V.~I.} \bibnamefont{Anisimov}},
  \bibinfo{author}{\bibfnamefont{F.}~\bibnamefont{Aryasetiawan}},
  \bibnamefont{and} \bibinfo{author}{\bibfnamefont{A.~I.}
  \bibnamefont{Lichtenstein}}, \bibinfo{journal}{Journal of Physics: Condensed
  Matter} \textbf{\bibinfo{volume}{9}}, \bibinfo{pages}{767}
  (\bibinfo{year}{1997}).

\bibitem[{\citenamefont{Becke}(1993)}]{hybrid}
\bibinfo{author}{\bibfnamefont{A.~D.} \bibnamefont{Becke}},
  \bibinfo{journal}{The Journal of Chemical Physics}
  \textbf{\bibinfo{volume}{98}}, \bibinfo{pages}{1372} (\bibinfo{year}{1993}).

\bibitem[{\citenamefont{Heyd et~al.}(2003)\citenamefont{Heyd, Scuseria, and
  Ernzerhof}}]{hse06:1}
\bibinfo{author}{\bibfnamefont{J.}~\bibnamefont{Heyd}},
  \bibinfo{author}{\bibfnamefont{G.~E.} \bibnamefont{Scuseria}},
  \bibnamefont{and}
  \bibinfo{author}{\bibfnamefont{M.}~\bibnamefont{Ernzerhof}},
  \bibinfo{journal}{The Journal of Chemical Physics}
  \textbf{\bibinfo{volume}{118}}, \bibinfo{pages}{8207} (\bibinfo{year}{2003}),
  ISSN \bibinfo{issn}{00219606}.

\bibitem[{\citenamefont{Heyd and Scuseria}(2004)}]{hse06:2}
\bibinfo{author}{\bibfnamefont{J.}~\bibnamefont{Heyd}} \bibnamefont{and}
  \bibinfo{author}{\bibfnamefont{G.~E.} \bibnamefont{Scuseria}},
  \bibinfo{journal}{The Journal of Chemical Physics}
  \textbf{\bibinfo{volume}{121}}, \bibinfo{pages}{1187} (\bibinfo{year}{2004}),
  ISSN \bibinfo{issn}{00219606}.

\bibitem[{\citenamefont{Heyd et~al.}(2006)\citenamefont{Heyd, Scuseria, and
  Ernzerhof}}]{hse06:3}
\bibinfo{author}{\bibfnamefont{J.}~\bibnamefont{Heyd}},
  \bibinfo{author}{\bibfnamefont{G.~E.} \bibnamefont{Scuseria}},
  \bibnamefont{and}
  \bibinfo{author}{\bibfnamefont{M.}~\bibnamefont{Ernzerhof}},
  \bibinfo{journal}{The Journal of Chemical Physics}
  \textbf{\bibinfo{volume}{124}}, \bibinfo{pages}{219906}
  (\bibinfo{year}{2006}), ISSN \bibinfo{issn}{00219606}.

\bibitem[{\citenamefont{Bl\"ochl}(1994)}]{bloch94}
\bibinfo{author}{\bibfnamefont{P.~E.} \bibnamefont{Bl\"ochl}},
  \bibinfo{journal}{Phys. Rev. B} \textbf{\bibinfo{volume}{50}},
  \bibinfo{pages}{17953} (\bibinfo{year}{1994}).

\bibitem[{\citenamefont{Kresse and Joubert}(1999)}]{kresse99}
\bibinfo{author}{\bibfnamefont{G.}~\bibnamefont{Kresse}} \bibnamefont{and}
  \bibinfo{author}{\bibfnamefont{D.}~\bibnamefont{Joubert}},
  \bibinfo{journal}{Phys. Rev. B} \textbf{\bibinfo{volume}{59}},
  \bibinfo{pages}{1758} (\bibinfo{year}{1999}).

\bibitem[{\citenamefont{Kresse and Hafner}(1993)}]{vasp1}
\bibinfo{author}{\bibfnamefont{G.}~\bibnamefont{Kresse}} \bibnamefont{and}
  \bibinfo{author}{\bibfnamefont{J.}~\bibnamefont{Hafner}},
  \bibinfo{journal}{Phys. Rev. B} \textbf{\bibinfo{volume}{47}},
  \bibinfo{pages}{558} (\bibinfo{year}{1993}).

\bibitem[{\citenamefont{Kresse and Furthm\"uller}(1996{\natexlab{a}})}]{vasp2}
\bibinfo{author}{\bibfnamefont{G.}~\bibnamefont{Kresse}} \bibnamefont{and}
  \bibinfo{author}{\bibfnamefont{J.}~\bibnamefont{Furthm\"uller}},
  \bibinfo{journal}{Computational Materials Science}
  \textbf{\bibinfo{volume}{6}}, \bibinfo{pages}{15 }
  (\bibinfo{year}{1996}{\natexlab{a}}).

\bibitem[{\citenamefont{Kresse and Furthm\"uller}(1996{\natexlab{b}})}]{vasp3}
\bibinfo{author}{\bibfnamefont{G.}~\bibnamefont{Kresse}} \bibnamefont{and}
  \bibinfo{author}{\bibfnamefont{J.}~\bibnamefont{Furthm\"uller}},
  \bibinfo{journal}{Phys. Rev. B} \textbf{\bibinfo{volume}{54}},
  \bibinfo{pages}{11169} (\bibinfo{year}{1996}{\natexlab{b}}).

\bibitem[{\citenamefont{Perdew et~al.}(1996)\citenamefont{Perdew, Burke, and
  Ernzerhof}}]{pbe}
\bibinfo{author}{\bibfnamefont{J.~P.} \bibnamefont{Perdew}},
  \bibinfo{author}{\bibfnamefont{K.}~\bibnamefont{Burke}}, \bibnamefont{and}
  \bibinfo{author}{\bibfnamefont{M.}~\bibnamefont{Ernzerhof}},
  \bibinfo{journal}{Phys. Rev. Lett.} \textbf{\bibinfo{volume}{77}},
  \bibinfo{pages}{3865} (\bibinfo{year}{1996}).

\bibitem[{\citenamefont{Dudarev et~al.}(1998)\citenamefont{Dudarev, Botton,
  Savrasov, Humphreys, and Sutton}}]{dudarev98}
\bibinfo{author}{\bibfnamefont{S.~L.} \bibnamefont{Dudarev}},
  \bibinfo{author}{\bibfnamefont{G.~A.} \bibnamefont{Botton}},
  \bibinfo{author}{\bibfnamefont{S.~Y.} \bibnamefont{Savrasov}},
  \bibinfo{author}{\bibfnamefont{C.~J.} \bibnamefont{Humphreys}},
  \bibnamefont{and} \bibinfo{author}{\bibfnamefont{A.~P.}
  \bibnamefont{Sutton}}, \bibinfo{journal}{Phys. Rev. B}
  \textbf{\bibinfo{volume}{57}}, \bibinfo{pages}{1505} (\bibinfo{year}{1998}).

\bibitem[{\citenamefont{Skoug et~al.}(2011)\citenamefont{Skoug, Cain,
  Majsztrik, Kirkham, Lara-Curzio, and Morelli}}]{Skoug.doping.2011}
\bibinfo{author}{\bibfnamefont{E.}~\bibnamefont{Skoug}},
  \bibinfo{author}{\bibfnamefont{J.}~\bibnamefont{Cain}},
  \bibinfo{author}{\bibfnamefont{P.}~\bibnamefont{Majsztrik}},
  \bibinfo{author}{\bibfnamefont{M.}~\bibnamefont{Kirkham}},
  \bibinfo{author}{\bibfnamefont{E.}~\bibnamefont{Lara-Curzio}},
  \bibnamefont{and} \bibinfo{author}{\bibfnamefont{D.}~\bibnamefont{Morelli}},
  \bibinfo{journal}{Science of Advanced Materials}
  \textbf{\bibinfo{volume}{3}}, \bibinfo{pages}{602} (\bibinfo{year}{2011}).

\bibitem[{\citenamefont{Nakanishi et~al.}(1969)\citenamefont{Nakanishi, Endo,
  and Irie}}]{nakanishi69}
\bibinfo{author}{\bibfnamefont{H.}~\bibnamefont{Nakanishi}},
  \bibinfo{author}{\bibfnamefont{S.}~\bibnamefont{Endo}}, \bibnamefont{and}
  \bibinfo{author}{\bibfnamefont{T.}~\bibnamefont{Irie}},
  \bibinfo{journal}{Jpn. J. Appl. Phys.} \textbf{\bibinfo{volume}{8}},
  \bibinfo{pages}{443} (\bibinfo{year}{1969}).

\bibitem[{\citenamefont{Yang et~al.}(2011)\citenamefont{Yang, Huang, Wu, and
  Xu}}]{yang11.newptype}
\bibinfo{author}{\bibfnamefont{C.}~\bibnamefont{Yang}},
  \bibinfo{author}{\bibfnamefont{F.}~\bibnamefont{Huang}},
  \bibinfo{author}{\bibfnamefont{L.}~\bibnamefont{Wu}}, \bibnamefont{and}
  \bibinfo{author}{\bibfnamefont{K.}~\bibnamefont{Xu}},
  \bibinfo{journal}{Journal of Physics D: Applied Physics}
  \textbf{\bibinfo{volume}{44}}, \bibinfo{pages}{295404}
  (\bibinfo{year}{2011}).

\bibitem[{\citenamefont{Blaha et~al.}(2005)\citenamefont{Blaha, Schwarz, and
  Novák}}]{Blaha.ldau.cu.2005}
\bibinfo{author}{\bibfnamefont{P.}~\bibnamefont{Blaha}},
  \bibinfo{author}{\bibfnamefont{K.}~\bibnamefont{Schwarz}}, \bibnamefont{and}
  \bibinfo{author}{\bibfnamefont{P.}~\bibnamefont{Novák}},
  \bibinfo{journal}{International Journal of Quantum Chemistry}
  \textbf{\bibinfo{volume}{101}}, \bibinfo{pages}{550} (\bibinfo{year}{2005}),
  ISSN \bibinfo{issn}{1097-461X}.

\end{thebibliography}

\end{document}